# Thermal Entanglement in Disordered Spin Chains: Localization, Thresholds, and the Quantum-to-Classical Crossover

Dihang Sun(孙頔杭) and Zhigang Hu(胡志刚)

*International Center for Quantum Materials, School of Physics, Peking University, Beijing 100871, China*

Biao Wu(吴飙)*

*International Center for Quantum Materials, School of Physics, Peking University, Beijing 100871, China*
*Wilczek Quantum Center, Shanghai Institute for Advanced Studies, Shanghai 201315, China and*
*Hefei National Laboratory, Hefei 230088, China*

(Dated: February 26, 2025)

We investigate the mixed-state entanglement between two spins embedded in the XXZ Heisenberg chain under thermal equilibrium. By deriving an analytical expression for the entanglement of two-spin thermal states and extending this analysis to larger spin chains, we demonstrate that mixed-state entanglement is profoundly shaped by both disorder and temperature. Our results reveal a sharp distinction between many-body localized (MBL) and ergodic phases, with entanglement vanishing above different finite temperature thresholds. Furthermore, by analyzing non-adjacent spins, we uncover an approximate exponential decay of entanglement with separation. This work advances the understanding of the quantum-to-classical transition by linking the entanglement properties of small subsystems to the broader thermal environment, offering a explanation for the absence of entanglement in macroscopic systems. These findings provide critical insights into quantum many-body physics, bridging concepts from thermalization, localization, and quantum information theory.

## I. INTRODUCTION

In the microscopic realm, quantum entanglement is a fundamental concept essential for understanding a wide array of quantum phenomena. It is pervasive in the Hilbert space of many-body systems, where a typical state is entangled. However, quantum entanglement seems conspicuously absent not only between macroscopic objects, such as cats that are composed of atoms and molecules, but also between microscopic particles in a thermal state, like atoms in the air. How can this absence be explained? A potential explanation lies in a recent result by Aubrun, Szarek, and Ye (ASY) [1]. They rigorously demonstrated that two subsystems are typically separable when both are coupled to a much larger third subsystem, provided the Hilbert space dimensions of the two subsystems are significantly smaller than that of the third subsystem. In our everyday experience, many objects are coupled to a common heat bath which is much larger, for instance, two atoms in the air or two cats in a room. Consequently, the ASY result is likely applicable to thermal states, suggesting that quantum entanglement is absent between two atoms in the air or two cats in a room. In this work we find that the ASY result indeed holds for many thermal states, providing a compelling explanation for the absence of quantum entanglement in our daily lives.

For simplicity, our study focuses on the entanglement between two spins, either in a thermal state or embedded within a larger spin chain. We employ concurrence to quantify the mixed-state entanglement between the two spins [2], and investigate how this entanglement is influenced by both temperature and the system's quantum phases—specifically, the many-body localization (MBL) phase versus the ergodic phase [3–10].

We begin by deriving an analytical expression for the entanglement of formation for thermal states of two spins with XXZ-type interactions, which serves as a benchmark for understanding more complex systems. Through numerical simulations [11, 12], we extend this analysis to longer spin chains, where the entanglement of a two-spin subsystem is studied by tracing out the rest of the chain. Our results reveal that the mixed-state entanglement is strongly dependent on both temperature and the underlying quantum phase. In the MBL phase, where disorder inhibits thermalization, the entanglement is generally lower compared to the ergodic phase. Moreover, at sufficiently high temperatures, the mixed-state entanglement vanishes entirely for both phases. For two non-adjacent spins in a chain at a given temperature, we find that the entanglement decreases roughly exponentially with separation then vanishes entirely for a larger separation. This obser-vation suggests that, for thermal states, quantum entanglement ceases to exist above a finite temperature threshold and for a larger separation, which is consistent with our experience in the macroscopic world.

---

* wubiao@pku.edu.cn

This paper is structured as follows: In Section II, we introduce the foundational concepts and mathematical framework for mixed-state entanglement, including the definitions and computational methods for concurrence and entanglement of formation. Section III presents an analytical solution for thermal entanglement in a two-spin system, serving as a key reference for our study. In Section IV, we detail our main results and discussions, focusing on the entanglement behavior in XXZ chains under various conditions. Special attention is given to the differences in thermal entanglement properties between the many-body localized (MBL) and ergodic phases. Finally, in Section V, we summarize our findings, discuss their implications for understanding quantum correlations in disordered and interacting systems, and suggest potential directions for future research.

## II. MIXED-STATE ENTANGLEMENT AND CONCURRENCE

When a bi-partite system is in pure state, the entanglement between its two subsystems, A and B, can be fully characterized using the well-known von Neumann entropy, defined as $E(\psi) = -\text{Tr}(\rho_A \log_2 \rho_A) = -\text{Tr}(\rho_B \log_2 \rho_B)$. Here, $\rho_A$ and $\rho_B$ are density matrices for subsystems A and B, respectively, obtained by partially tracing the pure state $|\psi\rangle\langle\psi|$.

However, for mixed states, the situation becomes more complex, and the von Neumann entropy can no longer directly quantify the entanglement between the two subsystems. In this case, the entanglement of formation (EoF) emerges as a more suitable measure. The EoF is defined as $E(\rho) = \min \sum_i p_i E(\psi_i)$, where $\rho = \sum_{i=1}^n p_i |\psi_i\rangle\langle\psi_i|$ is a mixed-state. Each $|\psi_i\rangle$ is a pure state of the bipartite system, with $0 \leq p_1, \ldots, p_n \leq 1$ and $\sum_{i=1}^n p_i = 1$. $E(\psi_i)$ is the von Neumann entropy of the pure state $|\psi_i\rangle$, and the minimization is taken over all possible decompositions of $\rho$ [13]. It is evident that if each $|\psi_i\rangle$ is a product state, the entanglement is zero, and $\rho$ is separable.

It is clear that from this definition that EoF is defined as the minimum amount of pure-state entanglement required to represent a mixed state. Unlike other popular entanglement monotones, such as the positive partial transpose (PPT) criterion, which only provides partial information of entanglement and is not a necessary and sufficient criterion, EoF provides a complete and robust characterization of mixed-state entanglement. While PPT is useful in certain scenarios, it is sometimes less reliable in distinguishing between separable and entangled states, particularly when the state is not close to a pure state. EoF, by contrast, is more accurate and consistent in such cases, as it directly reflects the entanglement resource needed to create a mixed state[13, 14].

In most cases, calculating $E(\rho)$ is not straightforward, as finding the minimum in its definition is numerically challenging. Fortunately, for a two-spin system, Wootters developed a practical method to compute it precisely [2]. The entanglement is given by

$$E(\rho) = h\left(\frac{1 + \sqrt{1 - C^2}}{2}\right) \tag{1}$$

where $h(x) = -x \log_2 x - (1-x) \log_2(1-x)$. Here, $C$ is the concurrence, defined as

$$C = \max\{0, \lambda_1 - \lambda_2 - \lambda_3 - \lambda_4\}, \tag{2}$$

and $\lambda_i$ are the square roots of the eigenvalues of the non-Hermitian matrix $\rho\tilde{\rho}$, arranged in decreasing order. The matrix $\tilde{\rho}$, known as the spin-flipped state, defined as $\tilde{\rho} = (\sigma_y \otimes \sigma_y)\rho^*(\sigma_y \otimes \sigma_y)$, where $\sigma_y$ is the standard Pauli matrix, and the complex conjugate * is taken in the standard basis. For a pair of spin-1/2 particles, the basis is $\{|\downarrow\downarrow\rangle, |\uparrow\downarrow\rangle, |\downarrow\uparrow\rangle, |\uparrow\uparrow\rangle\}$.

## III. TWO THERMAL SPINS

We consider two spins with an XXZ-type Heisenberg interaction in an inhomogeneous field. The Hamiltonian of the system is given by

$$H = J(S_1^x S_2^x + S_1^y S_2^y + \gamma S_1^z S_2^z) + h_1 S_1^z + h_2 S_2^z, \tag{3}$$

where $J$ is the coupling constant, representing the interaction strength between spins in the $x$ and $y$ directions, and $\gamma J$ represents the interaction strength in the $z$ direction. The parameters $h_1$ and $h_2$ denote the local magnetic fields acting on the two spins. For the thermal state $\rho = e^{-\beta H}/\text{tr}(e^{-\beta H})$, the concurrence can be derived analytically and is expressed as

$$C = \max(0, \chi),$$
$$\chi = \frac{\frac{1}{2}e^{\frac{\gamma\beta J}{2}}|\sinh(\xi\beta J)| - 1}{e^{\frac{\gamma\beta J}{2}}\cosh(\xi\beta J) + \cosh\left(\frac{(h_1+h_2)\beta}{2}\right)}, \tag{4}$$

where $\beta = 1/(kT)$ and $\xi = \sqrt{1+(\Delta h)^2}/2$ with $\Delta h = (h_1 - h_2)/J$. The concurrence vanishes when the temperature exceeds a threshold value $T_c$, which satisfies the equation

$$e^{\frac{\gamma\beta_c J}{2}}|\sinh \xi \beta_c J| = 2\xi \, , \qquad (5)$$

where $\beta_c = 1/(k_B T_c)$. This result is consistent with the ASY result [1], supporting the conjecture that entanglement between two systems in a thermal state vanishes at sufficiently high but finite temperatures. Rigorously proving this general result remains a challenging task, which we will address in future work.

It is also meaningful to consider the threshold in terms of $\beta\Delta E$, where $\Delta E = (E_{max} - E_{min})/3$, the average energy gap for the Hamiltonian given in Equation (3), is proportional to the average energy per spin. For a special case $h_1 + h_2 = 0$ and $\lambda < 1$, the threshold satisfies the equation

$$\exp\left[\frac{3\gamma}{4\xi}(\beta\Delta E)_c\right]\sinh[\frac{3}{2}(\beta\Delta E)_c] = 2\xi \, . \qquad (6)$$

In this case, we can rigorously prove that $\frac{\mathrm{d}(\beta\Delta E)_c}{\mathrm{d}\xi} > 0$, indicating that this threshold increases monotonically with $\Delta h$. This aligns with physical intuition, as a higher degree of entanglement in the ground state requires a greater temperature to completely eliminate the entanglement.

Let us consider a special case where $\gamma = 1$ and $h_1 = h_2$, corresponding to an XXX-type Heisenberg interaction with identical magnetic fields. This scenario has been studied previously in Refs. [15, 16]. When the interaction is ferromagnetic $J < 0$, the threshold equation (5) cannot be satisfied, implying $T_c = 0$, In other words, the thermal state is always separable, regardless of how low the temperature is. This is straightforward to understand, as the ground state for the two spins in this case is $|\uparrow\uparrow\rangle$, which is a product state. Conversely, when the interaction is antiferromagnetic $J > 0$, we find $T_c = J/(k_B \ln 3)$. This indicates that the two spins are entangled at sufficiently low temperatures, consistent with the fact that the ground state is entangled for $J > 0$, and a sufficiently high temperature is required to destroy this entanglement.

This simple example demonstrates that the entanglement present in the ground state persists in the thermal state at low temperatures. The general behavior of the entanglement for the two-spin thermal state is illustrated in Figure 1.

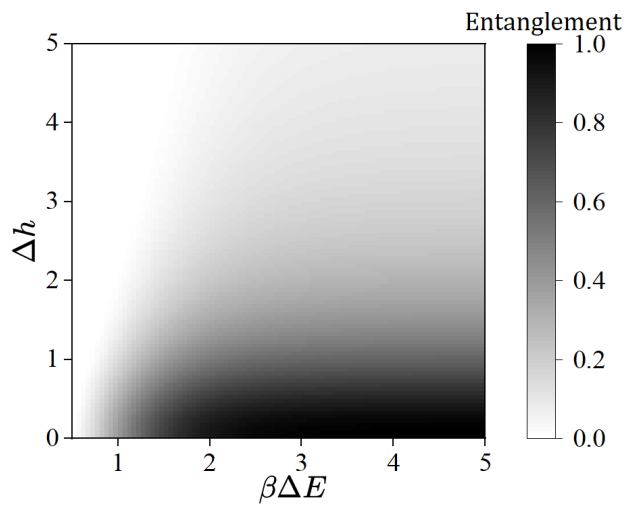

Figure 1: Entanglement of formation for a two-spin system. The gray intensity represents the magnitude of entanglement. The parameters are $\gamma = 0.4$ and $h_1 + h_2 = 0$.

## IV. TWO SPINS EMBEDDED IN CHAINS

Having gained a comprehensive understanding of entanglement properties in two-spin systems, it is natural to explore whether these insights can shed light on the entanglement characteristics of larger systems. For simplicity, we consider a relatively long XXZ spin chain and investigate the entanglement properties of a small subsystem embedded within this larger thermal system. By focusing on the induced states of such subsystems under varying conditions, we aim to understand the entanglement behaviors of small systems coupled to a common heat bath, which may reflect the situation for many macroscopic objects we encounter.

The Hamiltonian for the larger XXZ chain is given by:

$$H = J\sum_{i=1}^{L-1}(S_i^x S_{i+1}^x + S_i^y S_{i+1}^y + \gamma S_i^z S_{i+1}^z) + \lambda \sum_{i=1}^{L} h_i S_i^z \, , \qquad (7)$$

where $L$ stands for the length of the chain, $J$ is the coupling constant, $\gamma J$ represents the interaction strength in the $z$-direction, $\lambda$ represents the intensity of the disordered field, and $h_i$ are random values uniformly distributed between $-1$ and $1$.

We hope that the analytical results for the two-spin system can provide a foundation for studying larger sys-tems. A key question we address is how much the two-spin state differs from the induced state in the larger chain and how we can adjust the two-spin Hamiltonian to better approximate the induced state.

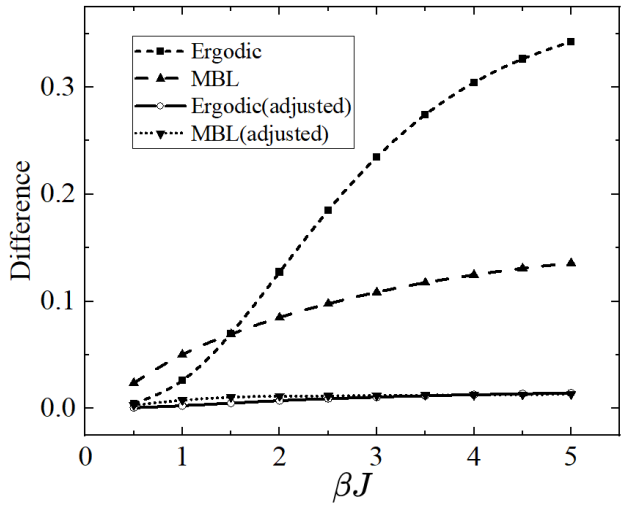

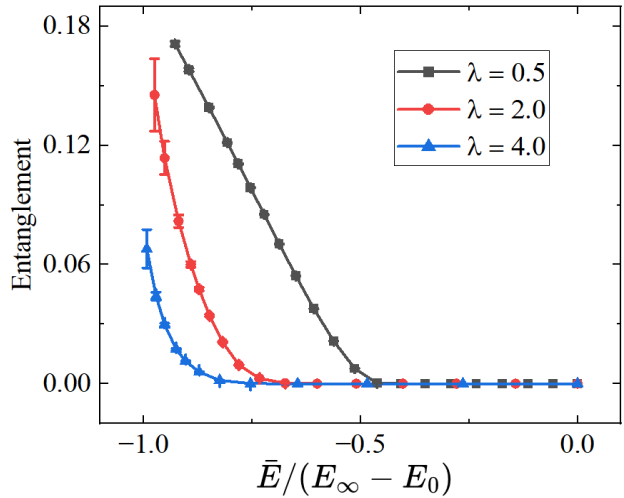

Figure 2: The short-dashed line with squares and the dashed line with triangles show the difference between the two-spin thermal state and the induced state from the large chain, using identical parameters ($\alpha_1 = \alpha_2 = 0$). The solid line with circles and the dotted line with triangles show the difference after adjusting the Hamiltonian using $\alpha_1$ and $\alpha_2$. The parameters chosen are $L = 100$, $\gamma = 0.4$, with the ergodic phase corresponding to $\lambda = 0.3$, and the MBL phase to $\lambda = 4.0$. These results demonstrate that the density matrix of the induced state from the large chain can be well approximated by the Hamiltonian given in Equation (8).

Figure 3: The entanglement of formation of the two-spin induced state for different conditions. $\bar{E}$ is the average energy for the two-spin state when regarded as thermal state of Hamiltonian given in Equation (8), $E_\infty$ is the thermal state energy at $T = \infty$, and $E_0$ represents the ground state energy. Due to the presence of random variables in the Hamiltonian, disorder averaging is performed to obtain statistically meaningful results, with error bars indicating the variance across multiple disorder realizations. The parameters chosen are $L = 100$ and $\gamma = 0.4$.

To achieve this, we define the Hamiltonian for the two-spin system as:

$$H_1 = (1 + \alpha_0)J\left(S_1^x S_2^x + S_1^y S_2^y + \gamma S_1^z S_2^z\right) + \lambda[(1 + \alpha_1)h_1 S_1^z + (1 + \alpha_2)h_2 S_2^z], \quad (8)$$

where $\alpha_0$, $\alpha_1$ and $\alpha_2$ are adjustable parameters. We assume that tracing out part of the spin chain can be approximated by modifying the local temperature at each site. Thus, $\alpha_1$ and $\alpha_2$ represent the influence of local temperature variations, while $\alpha_0$ accounts for the effect of local temperatures on the interaction term [17, 18]. For computational convenience, we approximate $\alpha_0$ with $(\alpha_1 + \alpha_2)/2$, reducing the number of adjustable parameters to two.

We consider the difference between two density matrices $\rho_1$ and $\rho_2$, where $\rho_1 = e^{-\beta H_1}/\text{tr}(e^{-\beta H_1})$ is the thermal state for the two-spin system, and $\rho_2$ is the density matrix for the two spins in the middle of the larger chain by tracing out the rest of the chain (referred to as the induced state), when the whole chain is in the thermal state $e^{-\beta H}/\text{tr}(e^{-\beta H})$. The parameters for $\rho_1$ and $\rho_2$ are exactly the same (including the coupling constant and disordered field). The difference is defined as $D = \frac{||\rho_1 - \rho_2||}{||\rho_1||}$, where $|| \cdot ||$ is the Frobenius norm.

To fit the induced state, we minimize the difference $D$ in the two-dimensional space of $\alpha_1$ and $\alpha_2$. We first perform a coarse global grid search to locate the minimum, followed by a gradient descent method with a smaller step size to refine the result. As shown in Figure 2, the difference between the two-spin state and the induced state becomes negligible when using the adjusted Hamiltonian, even at low temperature ($\beta J = 5.0$). This suggests that the entanglement behaviors of the two-spin induced state in larger spin chains are similar to those of the two-spin system, allowing the analytical results to be extended to larger systems.

The entanglement of formation and its variance are illustrated in Figure 3. Since the Hamiltonian contains random variables, we need to perform disorder averaging to obtain statistically meaningful results, with the variances arising from multiple disorder realizations. For $\lambda = 0.5$, the variance is nearly zero even at very low temperatures, indicating ergodic behavior. As $\lambda$ increases, the variance at low temperature grows, reflecting the onset of localization. The entanglement trends for the induced state closely resemble those of the two-spin system.

As expected, the quantum correlations in the thermal state of an MBL system are weaker than those in an ergodic system, because quantum states in a MBL system are highly affected by the disorder field on each site, unlike in an ergodic system[10, 19–21]. This is consistent with our observations in Figure 3. The energy threshold at $\lambda = 4.0$ is around $\bar{E} = -0.8(E_\infty - E_0)$, and as the disorder intensity decreases, the threshold increases. Here, $E_\infty$ is the thermal state energy at $T = \infty$, and $E_0$ represents the ground state energy. Note that the difference $E_\infty - E_0$ is proportional to the energy per spin.

The results in Figure 3 indicate that eliminating entanglement is more challenging in ergodic systems, though the induced state eventually becomes separable at sufficiently high temperatures. This is consistent with the findings in [22], which show the existence of a "separable ball" around the identity matrix, within which all mixed states are unentangled. As the temperature increases, the thermal state gradually becomes closer to the identity matrix, and eventually enters the "separable ball" at sufficiently high temperatures, making the bipartite entanglement disappear. This also suggests that for any physically sound quantum system, there exists a finite temperature threshold above which the entanglement between any of its two subsystems vanishes.

We now turn to non-adjacent spins. As shown in Figure 4, the mixed-state entanglement between two spins decays exponentially with increasing distance [23–25]. For spins separated by more than three sites, the mixed state is nearly separable. When the separation is five sites or more, the entanglement vanishes completely for all temperatures. This is primarily due to the nearest-neighbor interactions in our model, as entanglement between non-adjacent spins must propagate through intermediate spins. The influence of one spin on another diminishes rapidly with distance, consistent with the exponential decay of correlation functions in one-dimensional gapped systems [26]. A recent work by Bluhm *et al.* has shown that mutual information between distant regions in 1D quantum systems with local, finite-range interactions decays exponentially with distance, which further explains our results [27].

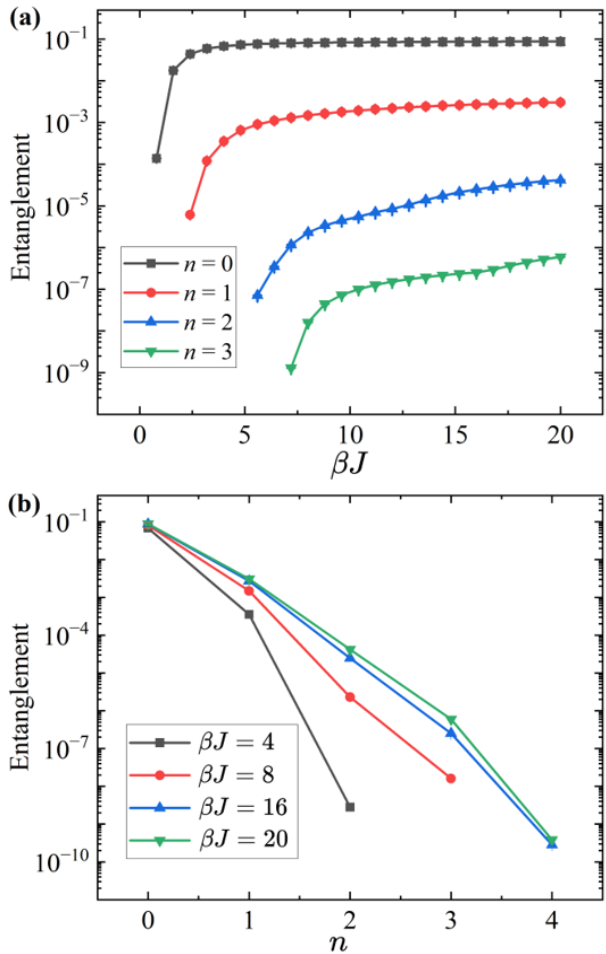

Figure 4: Entanglement for two non-adjacent spins. Here $n$ denotes the number of spins separating two non-adjacent spins, with $n = 0$ indicating adjacent spins. Our numerical calculations indicate that the entanglement between spins with separation $n \geq 5$ vanishes completely. The parameters chosen are $L = 100$, $\gamma = 0.4$, and $\lambda = 4.0$.

Furthermore, as shown in Figure 4 (b), the exponential decay of entanglement with distance becomes more pronounced at lower temperatures. At low temperatures, reduced thermal fluctuations allow correlation and thus entanglement to persist over longer distances, enhancing the exponential nature of the decay.

## V. CONCLUSIONS

In this study, we explored the mixed-state entanglement of thermal states in the XXZ Heisenberg chain, with a particular emphasis on distinguishing between the MBL phase and the ergodic phase. We derived an analytical expression for the entanglement of two-spin thermal states and utilized it to analyze spins embedded within larger spin chains. Our results reveal that the behavior of mixed-state entanglement is significantly influenced by both the disorder field and temperature, exhibiting a clear transition between localized and ergodic phases. We identified temperature thresholds above which the entanglement between two spins completely vanishes. Moreover, by extending our analysis to non-adjacent spins, we observed that entanglement in our model decays approximately exponentially with distance, with a sudden drop-off at larger distances. By examining the entanglement of two spins within larger thermal systems, this study provides partial insights into why quantum entanglement, which is ubiquitous in quantum many-body systems, is absent between both large objects (e.g., two cats) and small particles (e.g., two atoms in air) in our daily lives. In our classical world, it is feasible to define a local temperature for a small region, where the system can be considered to be in a thermal state. Our findings suggest that quantum entanglement between any two subsystems in such a locally thermal system disappears at sufficiently high temperatures or large distances. Furthermore, this work contributes to the broader research on many-body localization and thermalization by elucidating how varying conditions of disorder and interaction influence quantum correlations. These results not only enhance our understanding of mixed-state entanglement in quantum many-body systems but also pave new pathways for research in quantum phase transitions and quantum information theory.

## ACKNOWLEDGEMENTS

We thank Zhenhuan Liu and Chao Yin for stimulating discussions. This work was supported by the National Natural Science Foundation of China (Grants No. 92365202, No. 12475011, and No. 11921005), the National Key R&D Program of China (2024YFA1409002), and Shanghai Municipal Science and Technology Major Project (Grant No.2019SHZDZX01).

## Appendix: The analytical expression for the concurrence of two-spin thermal state

In this appendix, we will provide a detailed derivation of equation (4). We start from the Hamiltonian of the XXZ two-spin system

$$H = J(S_1^x S_2^x + S_1^y S_2^y + \gamma S_1^z S_2^z) + h_1 S_1^z + h_2 S_2^z \,,$$

for thermal state $\rho := e^{-\beta H}/\mathrm{tr}(e^{-\beta H})$, because of the symmetry constrains, the density matrix will be in this form:

$$\rho = \begin{pmatrix} u & 0 & 0 & 0 \\ 0 & w & z & 0 \\ 0 & z^* & w' & 0 \\ 0 & 0 & 0 & v \end{pmatrix} \,, \qquad (A.1)$$

where $u$, $v$ and $z$ can be expressed as

$$u = \frac{e^{-\frac{1}{2}(h_1 + h_2)}}{2[e^{\frac{\gamma\beta J}{2}}\cosh\xi\beta J + \cosh\frac{(h_1 + h_2)\beta}{2}]} \,,$$

$$v = \frac{e^{\frac{1}{2}(h_1 + h_2)}}{2[e^{\frac{\gamma\beta J}{2}}\cosh\xi\beta J + \cosh\frac{(h_1 + h_2)\beta}{2}]} \,,$$

$$z = \frac{e^{\frac{\gamma\beta J}{2}}\sinh\xi\beta J}{4\xi[e^{\frac{\gamma\beta J}{2}}\cosh\xi\beta J + \cosh\frac{(h_1 + h_2)\beta}{2}]} \,,$$

where $\xi = \sqrt{1 + (\Delta h)^2}/2$ and $\Delta h = (h_1 - h_2)/J$. Using the relation $C = \max\{0, \lambda_1 - \lambda_2 - \lambda_3 - \lambda_4\}$, we can find out that

$$C = 2\max(0, |z| - \sqrt{uv}) \,. \qquad (A.2)$$

Substitute the values of $u$, $v$ and $z$, the thermal state concurrence for XXZ model can be written as

$$C = \max\left(0, \frac{\frac{1}{2\xi}e^{\frac{\gamma\beta J}{2}}|\sinh\xi\beta J| - 1}{e^{\frac{\gamma\beta J}{2}}\cosh\xi\beta J + \cosh\frac{(h_1 + h_2)\beta}{2}}\right) \,, \qquad (A.3)$$

which is the analytical expression we provided in the main text.